\documentclass[10pt, conference, compsocconf]{IEEEtran}
\usepackage{multirow}
\usepackage{graphics}
\usepackage{pifont}
\usepackage{subfig}
\usepackage{footmisc}
\usepackage{threeparttable}
\usepackage{array}
\usepackage{color}

\ifCLASSINFOpdf
   \usepackage[pdftex]{graphicx}
\else
  \usepackage[dvips]{graphicx}
\fi

\usepackage[cmex10]{amsmath}
\begin{document}
\title{\textit{Ad hoc} Cloud Computing: From Concept to Realization}


\author{\IEEEauthorblockN{Gary A. McGilvary\IEEEauthorrefmark{1},
Adam Barker\IEEEauthorrefmark{2},
Malcolm Atkinson\IEEEauthorrefmark{1}}
\IEEEauthorblockA{\IEEEauthorrefmark{1}Edinburgh Data-Intensive Research Group, 
School of Informatics, The University of Edinburgh 
\\ Email: gary.mcgilvary@ed.ac.uk, mpa@staffmail.ed.ac.uk}
\IEEEauthorblockA{\IEEEauthorrefmark{2}School of Computer Science, University of St Andrews\\
Email: adam.barker@st-andrews.ac.uk}}

\maketitle

\begin{abstract}
This paper presents the first complete, integrated and end-to-end solution for \textit{ad hoc} cloud computing environments. \textit{Ad hoc} clouds harvest resources from existing sporadically available, non-exclusive (i.e. primarily used for some other purpose) and unreliable infrastructures. In this paper we discuss the problems \textit{ad hoc} cloud computing solves and outline our architecture which is based on BOINC.

\end{abstract}


\IEEEpeerreviewmaketitle

\section{Introduction}
This paper introduces and develops a prototype of an \textit{ad hoc} cloud computing framework. \textit{Ad hoc} clouds harvest resources from existing sporadically available, non-exclusive (i.e. primarily used for some other purpose) and unreliable infrastructures. Examples of such infrastructures range from personal infrastructure users with a number of underutilized computers, to startup companies through to large-scale organizational infrastructures. 

The nature of providing a cloud service by harvesting resources from a set of unreliable hosts does have similar elements to Grid and volunteer computing. Despite being similar to volunteer and Grid computing as well as clouds (e.g. Amazon EC2 \cite{amazon_ec2}) and clusters (e.g HTCondor \cite{Thain2005}), the \textit{ad hoc} cloud computing paradigm has many key differences. The \textit{ad hoc} cloud model:

\begin{itemize}
\item \textbf{Volunteer resources:} Operates over a set of non-exclusive and sporadically available hosts, which may be unpredictable in nature. This is in contrast to offering a service from a dedicated cloud, cluster or Grid infrastructure where each host's resources are fully committed to the service.
\item \textbf{Lack of trust:} Does not assume a level of trust exists between an end-user and the infrastructure provider; a relationship that currently exists between end-users, clouds, clusters and Grids.
\item \textbf{Ensures continuity:} Maintains service availability in the presence of host membership churn or failure to ensure job continuity when running over a set of unreliable hosts.
\item \textbf{Low interference:} Does not interfere with executing host processes, especially in cases where these important processes consume a varying amount of resources at any given time.
\item \textbf{Diverse workloads:} Targets a set of more diverse applications such as memory, I/O and disk-intensive tasks as opposed to typical CPU-intensive applications commonly executed by volunteer computing systems.
\end{itemize}

\noindent Our research has developed solutions to each of the challenges above, and to our knowledge no other research in this field has presented such a complete, integrated and end-to-end solution for \textit{ad hoc} cloud computing environments. In this paper we detail the research challenges and solutions of developing an \textit{ad hoc} cloud computing prototype. Primarily we focus on our cloud continuity solution, however we explore possible solutions for the remaining problems.

For simplicity and brevity in this paper, we assume that our implementation will be predominately deployed on Local Area Networks to ensure reasonable security and performance guarantees; we aim to extend it to Wide Area Networks after evaluating and optimizing the architecture. Subsequently, applications that require extremely high levels of security, may not be suited to the \textit{ad hoc} cloud, or perhaps any cloud implementation. Furthermore, interactive applications or those that write to external dependencies may not function as expected due to data inconsistencies when a host fails abruptly. Such applications may need further reliability mechanisms, however there are approaches to solve such problems \cite{Cully2008}.

The rest of this paper is organized as follows: Section 2 provides an overview of the concepts and foundations of \textit{ad hoc} cloud computing followed by an in-depth feature and implementation overview of our platform in Section 3. Section 4 briefly reports our initial evaluation while Section 5 outlines related research. Section 6 concludes with a summary and plans for future work.

\section{Concepts and Foundations} 
We now discuss the architecture, components and processes of the \textit{ad hoc} cloud computing platform. 

\subsection{System Overview}
An \textit{ad hoc} cloud harvests resources from existing non-exclusive and sporadically available hosts used by \textit{host users} (e.g. company employees) and exposes these resources to cloud jobs submitted by \textit{cloud users}. Cloud jobs are submitted to the \textit{ad hoc} server which then schedules jobs to \textit{ad hoc guests}, or virtual machines, running on each of the hosts within the cloud. Each host has an \textit{ad hoc client} installed to control the guest, monitor resources and provide state information to the server. In particular, within an \textit{ad hoc} cloud, we create a set of \textit{cloudlets}; a set of connected \textit{ad hoc} guests that offer a particular service or environment. Figure 1 gives a high-level overview of these concepts.

\begin{figure}[h!]
 \begin{center}
\includegraphics[width=0.45\textwidth]{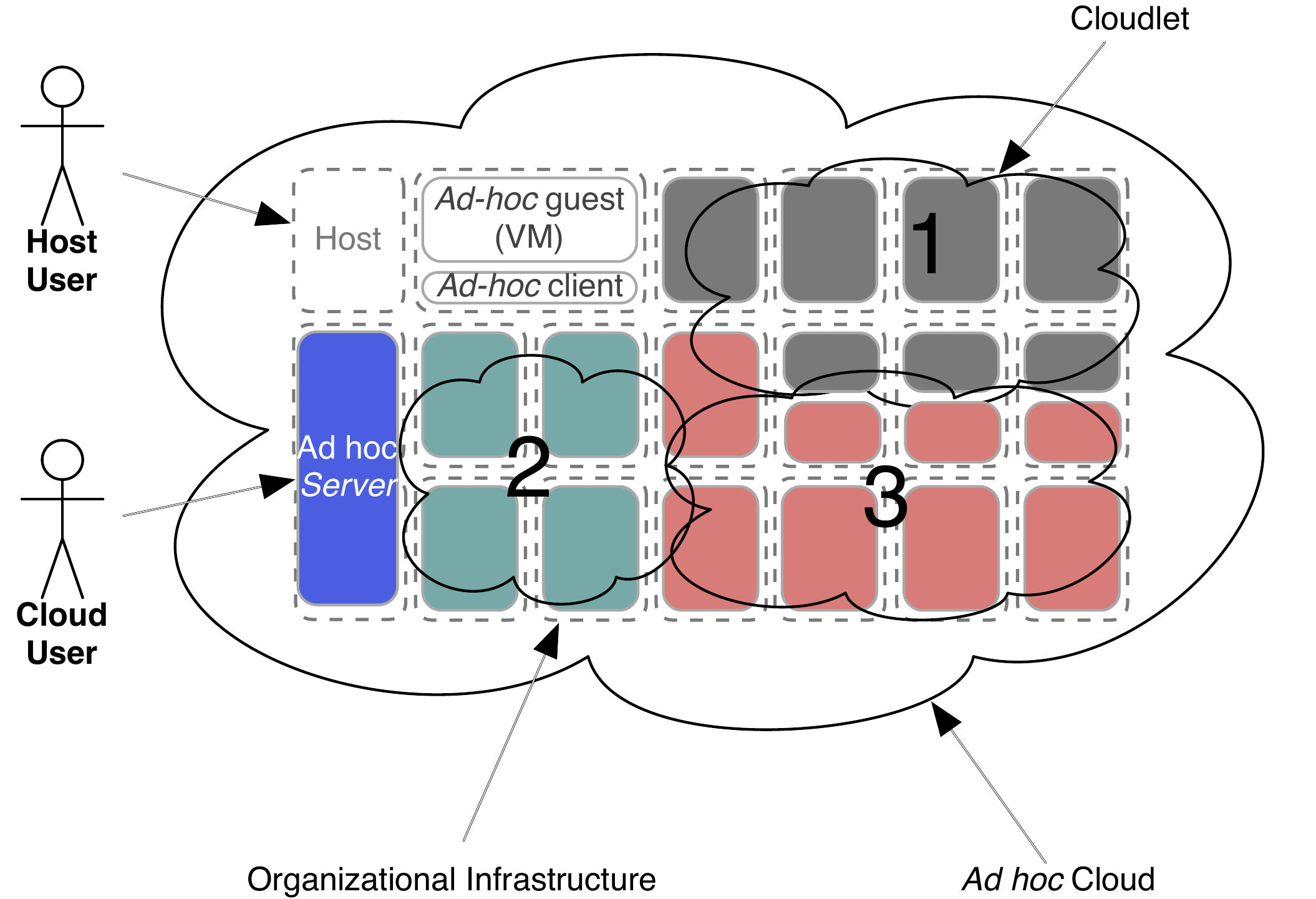}
 \end{center}
 \caption{The \textit{Ad hoc} Cloud Architecture}
\end{figure}

\noindent We see many cloudlets may exist which are deployed over multiple hosts to provide a unique service to the current infrastructure users. For instance, cloudlet `1'  and `2' may hold the necessary environments for Matlab and BLAST applications to execute respectively. A guest may either be dedicated to a single cloudlet, or part of many if multiple jobs are scheduled onto the same guest and require different operating environments. 

Grouping guests into cloudlets makes it easier to schedule jobs and monitor and manage the infrastructure. For example, only hosts within a specific cloudlet need to be taken into account when scheduling a job destined for that cloudlet. Cloudlets can also be managed to determine if a cloudlet's resources should be adjusted based on current computational requirements. Note that for simplicity, cloudlet members are co-located in Figure 1, however in reality these will be physically distributed across an infrastructure.

Cloudlets, hosts and guests are all centrally managed via the \textit{ad hoc} server, for example, the server accepts jobs, schedules jobs to hosts, maintains system state and ensures cloud jobs successfully complete. From the perspective of a cloud user, the processes and resources that underpin the \textit{ad hoc} cloud are hidden, therefore our cloud implementation conforms to a Platform as a Service (PaaS) infrastructure. With minor adjustments, such as passing the virtual machine's IP address back to the cloud user, the \textit{ad hoc} cloud can be converted into an Infrastructure as a Service (IaaS) cloud, giving cloud users the ability to login and control their assigned virtual machines. 

We now give an outline of the events that occur from cloud user job submission to result retrieval; these processes are explained further in subsequent sections.

\begin{itemize}
\item A cloud user submits a job (application and/or data) to a job submission service on the \textit{ad hoc} cloud server via a web interface.
\item A host which is instructed or wishes to donate their machine to the \textit{ad hoc} platform first connects to the \textit{ad hoc} server. A virtual machine is sent to their machine and the host then awaits further instructions.
\item The \textit{ad hoc} server schedules a job to a host for execution based on host reliability. Upon a job arriving, the host starts the virtual machine and job execution begins.
\item The \textit{ad hoc} client dynamically controls the virtual machine resources to minimize cloud job interference with executing host processes. Furthermore the client monitors the resource loads of both host and guest and passes this information to the server on a regular basis.
\item Periodically the \textit{ad hoc} client informs the \textit{ad hoc} server of host and virtual machine availability to indicate that they have not failed. 
\item Periodically the \textit{ad hoc} client captures virtual machine snapshots and distributes these in a Peer-to-Peer (P2P) fashion to other reliable and available hosts within the platform; only the most recent snapshot is stored to minimize disk consumption.
\item Upon a host or residing guest failing, the \textit{ad hoc} server detects this and restores the guest's most current snapshot upon another host to allow the job to continue. 
\end{itemize}

\subsection{Integrating Volunteer Computing and Virtualization}
The foundation of our \textit{ad hoc} platform relies on BOINC; an open source client-server volunteer computing middleware system \cite{Anderson2004}. As the \textit{ad hoc} cloud computing paradigm also involves harvesting resources from unreliable hosts, BOINC is well suited to our purposes. However, in order to protect cloud jobs running upon unreliable and perhaps unsecure hosts, as well as the need to protect the processes running on the hosts themselves, we have created a virtualized version of BOINC, named V-BOINC \cite{McGilvary2013}. V-BOINC takes advantage of virtualization to run BOINC computations within virtual machines as opposed to directly on volunteer machines.

V-BOINC sends these lightweight VirtualBox virtual machines from a V-BOINC Server to volunteer BOINC hosts when instructed by the host's installed modified BOINC client, called the V-BOINC Client. A regular BOINC client within the virtual machine is used to download a BOINC project's task. The process of obtaining a virtual machine and a BOINC task is shown in Figure 2.

\begin{figure}[h!]
  \begin{center}
\includegraphics[width=0.5\textwidth]{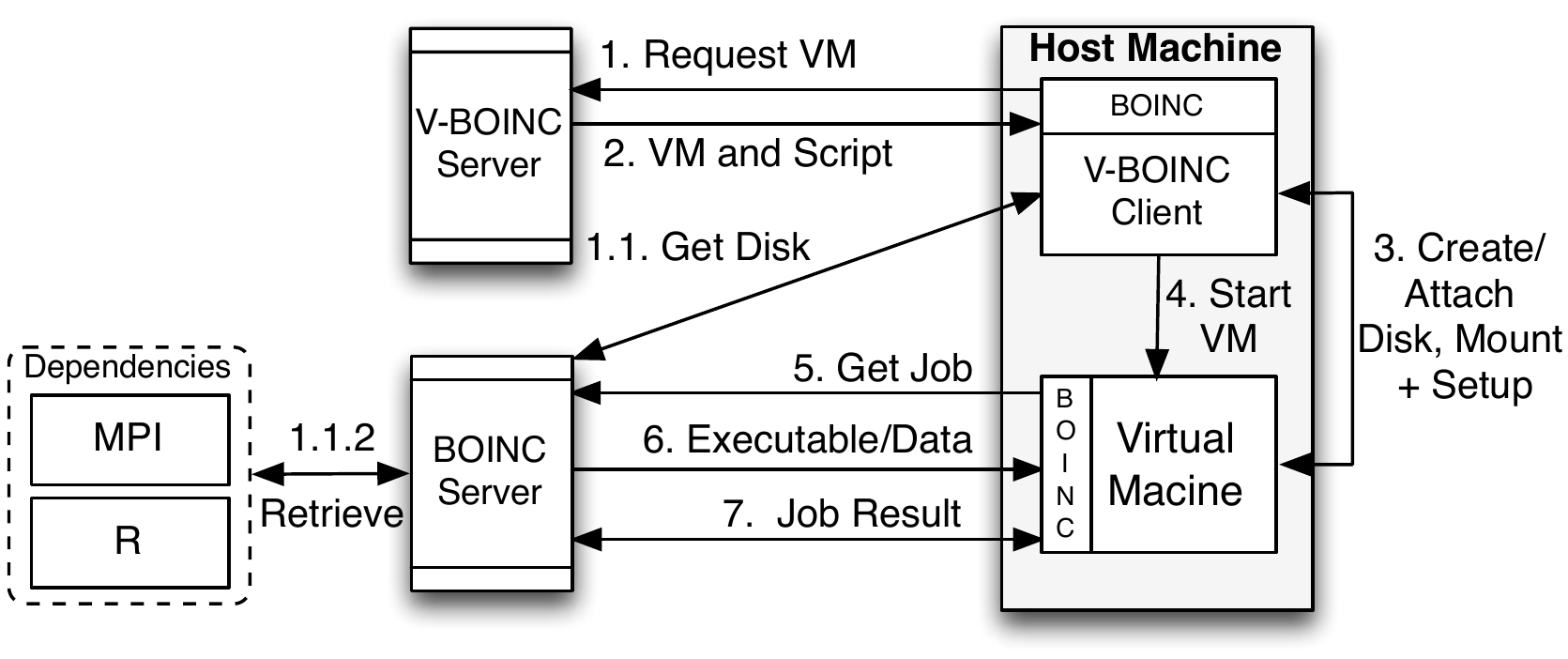}
  \end{center}
 \caption{V-BOINC Overview}
\end{figure}

Upon a volunteer user submitting the details of both the regular BOINC scientific project and the V-BOINC project they wish to attach to via the V-BOINC Client, the host is first instructed to request a virtual machine image (1) from the V-BOINC project. Concurrently, the V-BOINC Client probes the regular BOINC server to determine if any dependencies exist for the specified project (1.1); which can be downloaded and attached to the virtual machine. This gives the \textit{ad hoc} cloud the ability to execute a set of diverse workloads that have dependencies (e.g MPI, data sets, etc), however for the context of this paper, we omit these details.

The V-BOINC Server sends the virtual machine image and a script that configures it (e.g. sets CPUs, memory and disk space limits) to the V-BOINC Client (2). The virtual machine is then configured (3) and started (4) to allow it to request (5), receive (6) BOINC jobs and return job results (7). More information about V-BOINC can be found in \cite{McGilvary2013}.

\section{From Volunteer to Cloud Computing}
We now describe how the \textit{ad hoc} cloud was developed and the major components that underpins the concept. The basis of this work involves transforming V-BOINC into a platform that not only takes into account the requirements of a cloud computing system but also one that can operate over an unreliable infrastructure. 

As such, all subsequent modifications and features mentioned have been made either to the V-BOINC Server or V-BOINC Client, which are now named the \textit{ad hoc} server and \textit{ad hoc} client respectively. To ensure the \textit{ad hoc} server does not become a single point of failure, the server can be replicated and load balanced in the same way regular BOINC servers currently are. We also assume that an \textit{ad hoc} client and virtual machine both reside on each host within the \textit{ad hoc} infrastructure, i.e. processes (1) and (2) of Figure 2 have been completed.

\subsection{BOINC Job Submission}
To utilize the resources available in the \textit{ad hoc} cloud, a user must submit a job to the service; we currently assume a job is an application executable with an option to upload data to be analysed. In the case of V-BOINC, a virtual machine obtains an application by connecting to a specific BOINC project; BOINC project jobs are however statically created before the BOINC service begins. In contrast, we would like cloud users to submit any job at any moment in time, hence the \textit{ad hoc} cloud must allow user-defined jobs to be submitted on-the-fly to BOINC while the service is running. This is not a trivial task and other research has taken place to enable job submission to BOINC (e.g. \cite{Rios2011,Urbah2009,Kertesz2010}) however these methods either split single tasks into independent jobs to be executed or would generate too much overhead for this type of platform.

To enable on-the-fly and independent job submission to BOINC, we created a BOINC project named \textit{Job Service} upon the \textit{ad hoc} server to accept and distribute jobs. This is in addition to the V-BOINC project, which is modified and renamed to \textit{VM Service}, that allows hosts within the \textit{ad hoc} platform to obtain virtual machines. Figure 3 shows how both the job and virtual machine BOINC projects interact.

\begin{figure}[h!]
  \begin{center}
\includegraphics[width=0.3\textwidth]{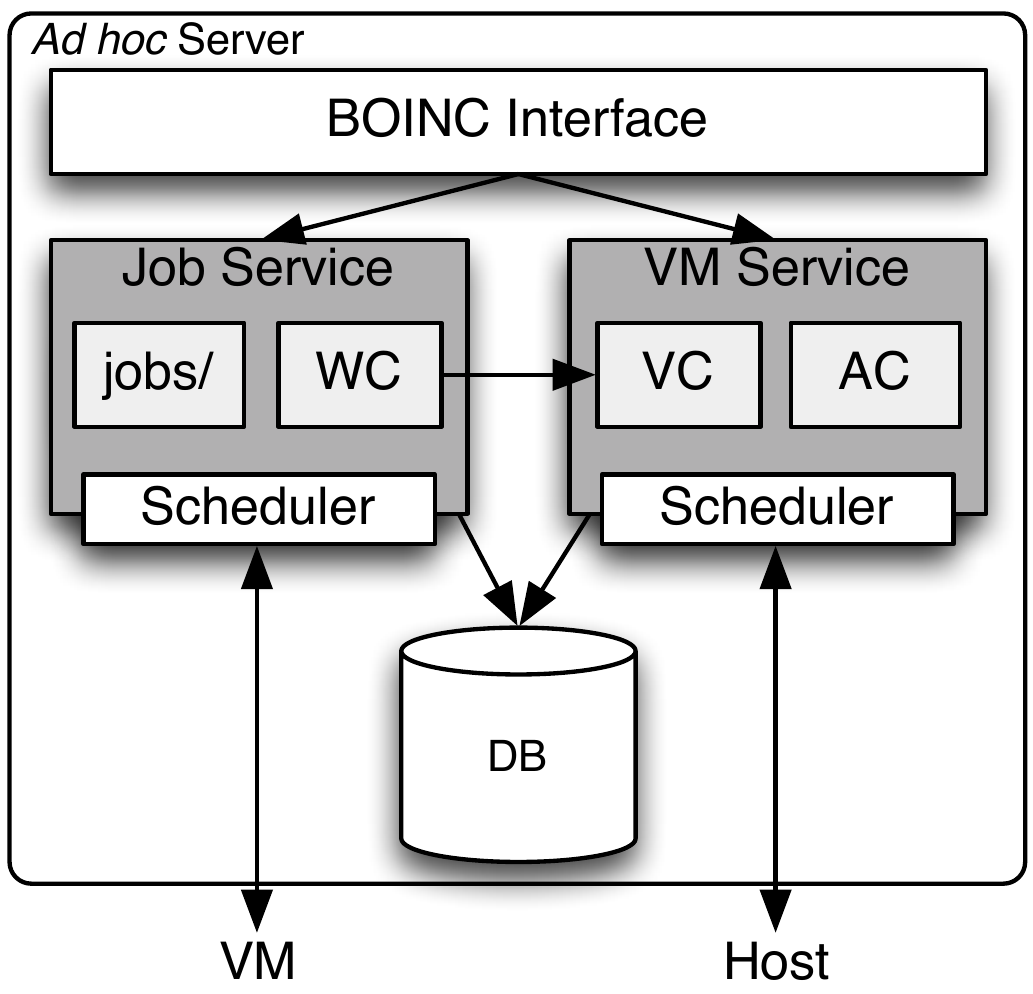}
  \end{center}
 \caption{The \textit{Ad hoc} Server Architecture}
\end{figure}

\noindent First a cloud user uploads an application and optional data via a web interface. All uploaded files to the service are placed in a \textit{jobs/} folder within the \textit{Job Service} project which are then processed by our developed \textit{work\_creator} (WC) daemon; BOINC allows project developers to create and add daemons to the BOINC default daemon set to perform user-defined actions. The \textit{work\_creator} daemon distinguishes application from data, creates XML descriptions of these files and calls BOINC API functions to ultimately create a BOINC workunit.

The \textit{Job Service} then informs the \textit{VM Service} that a cloud job exists and that the \textit{VM Service}'s \textit{vm\_controller} (VC) daemon can begin instantiating a virtual machine upon a volunteer host to execute this job. Currently failed volunteer hosts are determined by the \textit{availability\_checker} (AC) daemon which deems a host failed if they do not poll within two minutes; a host is set to periodically poll the server every minute.

\subsection{Reliability Scheduling}
Upon being notified a job awaits for execution, the \textit{VM Service} begins scheduling a job to the most reliable host with a virtual machine  ready to be used. The scheduler does this based on the following characteristics for each host: 

\begin{enumerate}
\item The total number of cloud jobs previously assigned,
\item The total number of cloud jobs previously completed,
\item The number of host failures, e.g. host termination, hardware or OS failures,
\item The number of guest failures, e.g. virtual machine configuration, instantiation, execution, and shutdown errors,
\item The current resource load.
\end{enumerate}
 
\noindent The reliability factors (1)-(3) are monitored by the \textit{ad hoc} server and are recorded in the Job Service database. The number of \textit{ad hoc} host failures are monitored by the VM Service's \textit{availability checker} daemon which sets an \textit{ad hoc} host to terminated or failed after two minutes of inactivity. 

The reliability factors (4)-(5) are monitored by the \textit{ad hoc} client when a cloud job is assigned to execute. To determine whether an \textit{ad hoc} guest is still executing, it is periodically polled every ten seconds using VirtualBox's \textit{VBoxManage} API ensuring those that have failed are detected quickly with minimal resource overheads. Any unexpected behaviour is reported to the server. The current resource loads of an \textit{ad hoc} host can be monitored either by Ganglia or via BOINC's basic monitoring mechanisms. For brevity, we omit the incorporation of resource load within our reliability scheduling due the significant size of the topic. More information about our reliability scheduling can be found at \cite{McGilvary2014}.

Based on the data sent from the \textit{ad hoc} client and the data collected on the \textit{ad hoc} server, the server is then able to calculate the reliability of each \textit{ad hoc} host using the following formula:

\[ host\_reliability = \left\{ 
  \begin{array}{l l}
     0 & \quad \text{if $NF = CA$}\\
    100 & \quad \text{if $NF = 0$}\\
    (CC/CA)*100 & \quad \text{otherwise}
  \end{array} \right.\]
\noindent where,
\begin{center}$NF$ = the total number of \textit{ad hoc} host and guest failures,\\
$CA$ = the total number of cloud jobs assigned to the host,\\
$CC$ = the total number of cloud jobs completed by the host.
\end{center}

\noindent An \textit{ad hoc} host's reliability is calculated after a cloud job has completed, when the \textit{ad hoc} guest has become non-operational or when the \textit{ad hoc} host has not polled within the last two minutes. The calculated reliabilities are then stored in the VM Service project database alongside the information of each candidate \textit{ad hoc} host. This reliability calculation gives an estimate of the \textit{ad hoc} host's behaviour for the entire time the host is part of the \textit{ad hoc} cloud.

\subsection{Transferring Control}
As a volunteer platform, BOINC and its virtual counterpart V-BOINC, are inherently controlled by the volunteer users. If a host user wishes to donate resources to a scientific project, receive jobs, change the amount of resources used or detach from a project, a host user is free to do so. However within an \textit{ad hoc} cloud platform, the server has to instruct the \textit{ad hoc} client to perform tasks, in turn transforming a client controlled infrastructure into a sever controlled infrastructure; we have modified our V-BOINC platform to allow this. The \textit{ad hoc} client receives and issues these commands via the \textit{ad hoc} BOINC client component to other middleware components shown in Figure 4. 

\begin{figure}[h!]
  \begin{center}
\includegraphics[width=0.42\textwidth]{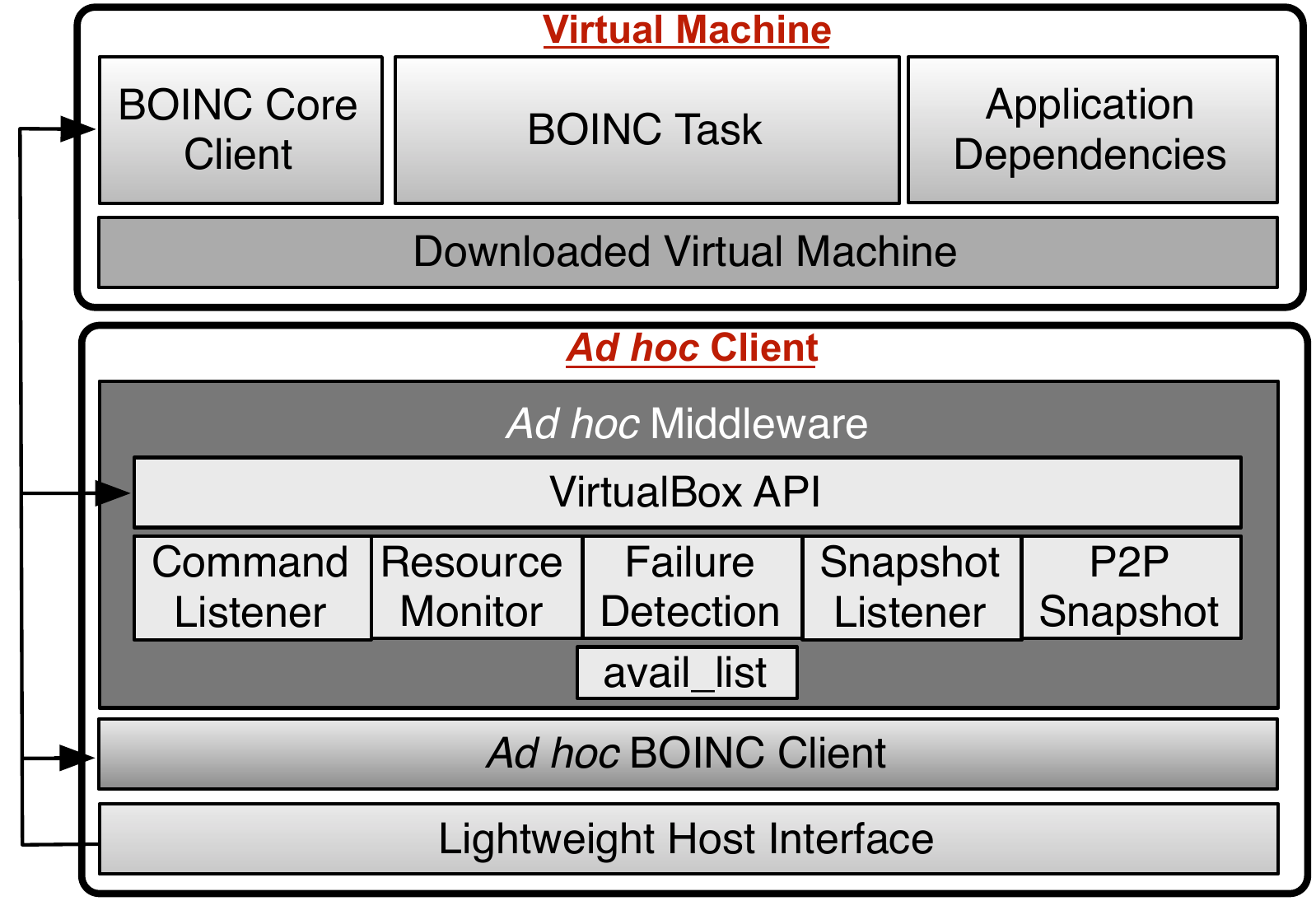}
  \end{center}
 \caption{The \textit{Ad hoc} Client Architecture}
\end{figure}

\noindent After the scheduler decides which virtual machine is to execute the cloud job, the server's \textit{VM Service} instructs the \textit{ad hoc} client to start the virtual machine on the host. It does so by passing the command from the \textit{vm\_controller} to the client's Command Listener via a BOINC XML message. The server also passes a description of the job such as whether any dependencies exist and the URL to download the correct job. If the former applies, the \textit{ad hoc} client downloads the dependencies and attaches it to the virtual machine before it begins, otherwise the virtual machine simply begins and it is then instructed to attach to the \textit{Job Service} to obtain a job. Commands relating to the operation of virtual machines are passed to the VirtualBox API for execution.

Cloud users are able to execute such commands via the Interface, which in turn informs the server to instruct the \textit{ad hoc} client to perform virtual machine commands via the VirtualBox API. Similarly, host users also have access to this and the regular commands of BOINC via our modified BOINC API. 

Figure 4 also shows the Resource Monitor and Failure Detection components. The Resource Monitor uses BOINC's default monitoring daemon and sends information back to the server for use in scheduling decisions. The component also monitors the current resource usage of the BOINC task to ensure the resource limits specified by the host user are not exceeded for prolonged periods of time, in which case the \textit{ad hoc} client will suspend the virtual machine until resource utilization has dropped to an acceptable level. This ensures that there is minimal interference with operating host processes.

In order to ensure the system state is up-to-date at any given time, each host periodically polls the \textit{ad hoc} server to signify the host's and virtual machine's availability. The latter is tested via the Failure Detection component; an independent process to test virtual machine availability. In response to a host poll, the \textit{ad hoc} server returns a list of all other available hosts in the BOINC XML message along with their IP addresses and reliability values.

\subsection{Making the Unreliable Reliable}
These reliability values (discussed in the previous Section) are used by our P2P Snapshot component, shown in Figure 4, which has the task of periodically taking virtual machine snapshots and transferring these to other clients in parallel using \textit{pssh} \cite{pssh} to ensure cloud job continuity. After a successful transfer, the \textit{ad hoc} server is informed of receiving hosts which now store the snapshot the location(s). When a host or virtual machine fails, the server is then able to instruct one of the receiving hosts to restore the snapshot. Take Figure 5 as an example, where each host's percentage failure probability is displayed.

\begin{figure}[h!]
  \begin{center}
\includegraphics[width=0.39\textwidth]{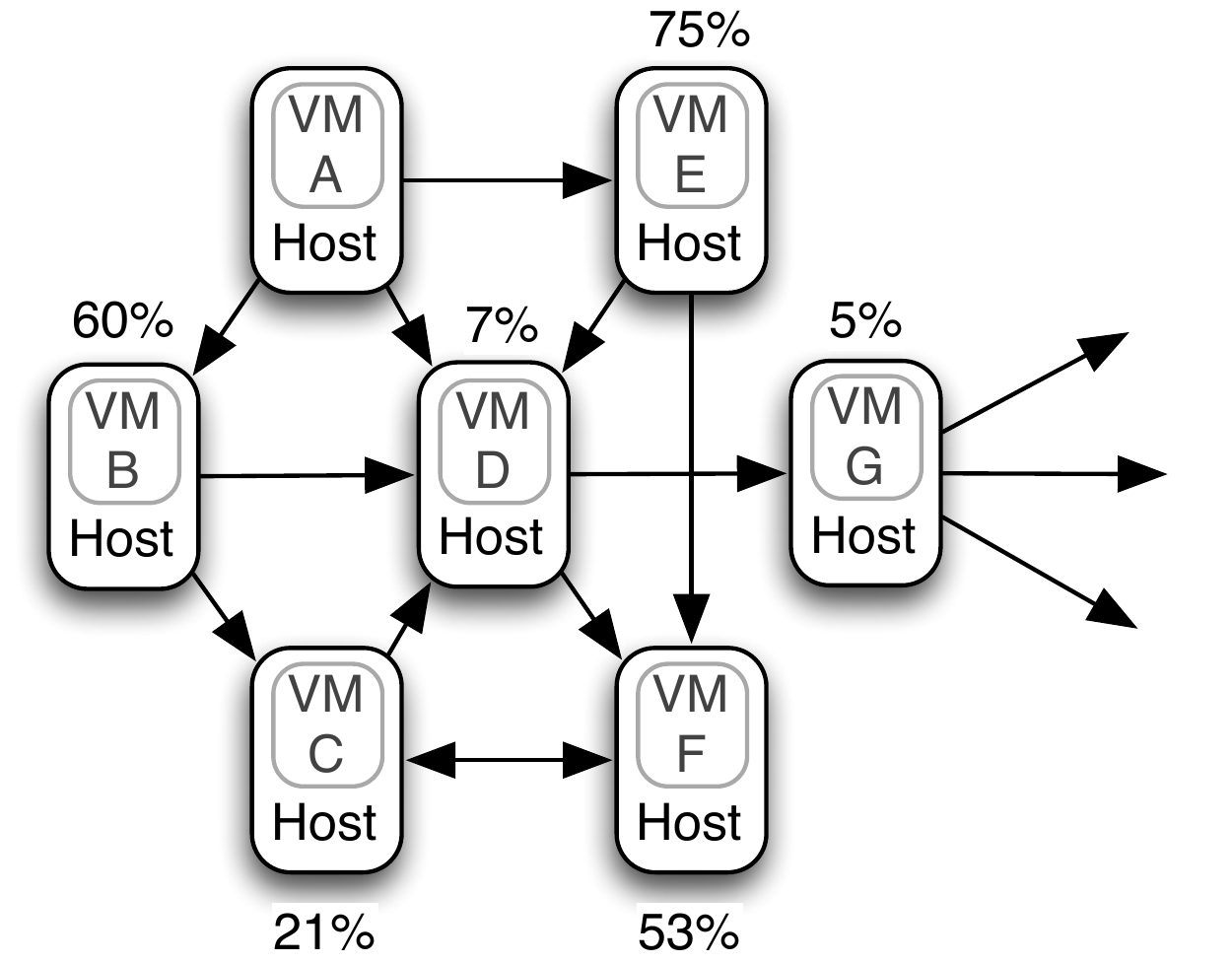}
  \end{center}
\caption{Snapshot Scheduling and Failure Probabilities (\%).}
\end{figure}

\noindent We have seven nodes A to G. At time \textit{t=0} node A checkpoints the virtual machine and the snapshot is sent to nodes B, D and E at \textit{t=1}. The virtual machine A terminates due to a host failure (\textit{t=2}), the \textit{ad hoc} server detects this (\textit{t=3}) and instructs node D to restore virtual machine A's snapshot (\textit{t=4}). This process is repeated for each node where periodic snapshots are taken and sent to others. Snapshots are however not blindly pushed to a random subset of hosts.

The P2P Snapshot component schedules snapshots to be stored on the most reliable hosts. It does this by filtering potential snapshot receivers based on whether they are in use, the sender's cloudlet membership and the descending reliability of potential receivers, all of which are taken from the list of available hosts within the BOINC XML message sent from the server. The algorithm will then select the first \textit{n} hosts that have less than a 5\% chance of all \textit{n} failing.

To maintain a reliable service, we want to ensure that 95\% of the time a cloud job will complete successfully. To satisfy this requirement, we must always have at least one current snapshot present on another host as its presence is directly related to the future success of an application if it is interrupted in any way. As such, we require that the combined probability of a group of hosts failing that store a particular virtual machine's snapshot is less than or equal to 5\%; this can be calculated by multiplying the respective failure probabilities of each host. For example, Figure 5 shows that the probability of a cloud job never completing while running on virtual machine A is 0.03\%. 

This scheduling method does however mean that reliable destinations may end up storing many snapshots. However, the maximum host storage that can be used by the \textit{ad hoc} cloud (e.g. the \textit{ad hoc} client, snapshots, etc) can be specified by the \textit{ad hoc} host user via regular BOINC. In the event this limit is reached, the \textit{ad hoc} server does not send the details of that host to polling \textit{ad hoc} clients, ensuring further snapshots are not sent to the host. We note there are many improvements that could be made to our P2P Snapshot Scheduler, however we leave these for future work.

In the event a virtual machine or host running a cloud job is deemed unreachable, the server begins the process of restoring the virtual machine's snapshot on another host. It does this based on the \textit{host\_reliability} formula when selecting the best host for initially deploying a job onto a virtual machine. Finally, all hosts that store the restored snapshot are instructed to delete it.

\section{Evaluation}
The prototype of our \textit{ad hoc} cloud computing concept was evaluated in terms of reliability and performance. Our reliability experiment tested our prototype running on 30 nodes the EDIM1 cluster \cite{Martin2011}. In order to accurately simulate an unreliable infrastructure, we obtained Nagios monitoring data over a period of 36 months from 650 hosts in The School of Informatics at The University of Edinburgh. We parsed this monitoring data, calculated the host activity for every hour and selected the hour where 30 hosts had the most activity.

We replayed these events on EDIM1 with our \textit{ad hoc} cloud installed and measured the completion rates of a variety of workloads; our prototype achieved up to 93.3\% reliability. This is a significant improvement to BOINC, which simply restarts a failed task, unless application checkpointing is enabled, in turn requiring application modification and the checkpoint to be present in memory.

In order to evaluate performance, we compared the times of executing a cloud job on our \textit{ad hoc} cloud and Amazon EC2 \cite{amazon_ec2}) instance with similar resources. We showed that our \textit{ad hoc} cloud can offer similar performance for a variety of workloads, even in the event of one or multiple \textit{ad hoc} guest failures, when taking into account the various overheads of both models. More information about our evaluation of the \textit{ad hoc} cloud can be found in \cite{McGilvary2014}.

\section{Related Work}
Authors of \cite{Kirby2010} propose the concept of the \textit{ad hoc} cloud within enterprise settings to harness unused resources to improve overall utilization, reduce net energy consumption and allow organizations to take advantage of operating their own in-house cloud. Their focus of the paper is to outline the major implementation challenges and describe one approach to creating an \textit{ad hoc} cloud computing infrastructure. The main challenges outlined relate to coping with the sporadically available hosts and how to minimize the impact on non-cloud processes to an acceptable level. 

Chandra \textit{et al.} propose a similar idea using Nebulas (synonymous to an \textit{ad hoc} cloud) where volunteer resources are used to create a cloud platform \cite{Chandra2009}. They note that Nebulas are particularly useful for applications that do not have strong performance guarantees and hence the authors focus on the performance and reliability of such platforms.

Sundarrajan \textit{et al.} describe their early experience with a prototype of their Nebula cloud system \cite{Weissman2011}; this was tested using a data-intensive blog analysis use case on PlanetLab \cite{planetlab}. Their architecture consists of the same core components found in our system, namely a master server to coordinate all activities backed by a database and clients to execute cloud jobs. The authors do not use virtual machines to execute cloud jobs but instead use the NativeClient plugin for web browsers to execute code as web applications. Their method shows that at times, little overhead can be seen when executing code in this way.

\section{Conclusions}
We have outlined our developed \textit{ad hoc} cloud computing platform that deploys a cloud service upon an end-user's existing infrastructure where member hosts are sporadically available and used for some other primary purpose. The \textit{ad hoc} cloud concept is useful for those who wish to improve their infrastructure efficiency and utilization as well as reduce costs by improving their return on IT investments. Furthermore, those who are not able to or do not wish to to migrate to the commercial or private cloud models can experiment and explore the potential of \textit{ad hoc} clouds before adopting either of the commercial or private models. 

In order to successfully develop an \textit{ad hoc} cloud computing platform, a large number of technological and research fields must be visited such as virtualization, volunteer computing and scheduling. Despite this, we showed by outlining our architecture that the concept of \textit{ad hoc} cloud computing is feasible and based on our initial evaluation, can be reliable and offer comparable performance to Amazon EC2. We are confident that the reliability and performance of our \textit{ad hoc} cloud development can increase in a range of other scenarios, however these extrapolations are based on assumptions that need to be tested by deploying the \textit{ad hoc} cloud on a live operational infrastructure with real workloads in the near future. A more detailed insight into our \textit{ad hoc} cloud computing prototype, the challenges solved and performance can be found at \cite{McGilvary2014}.

\bibliographystyle{IEEEtran}	
\bibliography{Papers_Read,Other_Refs,My_Papers}

\end{document}